\documentstyle[tighten,eqsecnum,aps,prd]{revtex}

\def\be{\begin{equation}}
\def\ee{\end{equation}}

\def\t{\theta}

\begin{document}

\title{Nariai metric is the first example of the singularity free model} 
\author{Naresh Dadhich\footnote{Electronic address: {\em nkd@iucaa.ernet.in
}}}
\address{Inter-University Centre for Astronomy and Astrophysics, Post 
Bag 4,Ganeshkhind,\\ Pune 411007, India}


\maketitle
\begin{abstract}
This is just to point out that the Nariai metric is the first example of the 
singularity free expanding perfect fluid cosmological model satisfying the 
weak energy condition, $\rho>0, \rho+p=0$. It is a conformally non-flat
Einstein space.
\end{abstract}

\vspace{0.5cm}

 Following the all powerful singualrity theorems [1] and the discovery of
the cosmic microwave background radiation [2] in mid sixties, the
existence of singularity in the relativistic cosmological models was without 
question taken for granted. This view was first challenged by Senovilla's
discovery [3]  in 1990 of the singularity free cylindrically symmetric
radiation model. The theorems did not apply in this case as the solution
did not
satisfy the assumption of occurrence of compact trapped surface. It is
noteworthy that violation of this assumption entailed no unphysical
features. This assumption seriously compromises, as is demonstrated by
this example, the generality of the theorems. It has always been looked
upon as suspect. From the physical point of view, this
assumption could be quite acceptable for gravitational collapse but
certainly not so for cosmology.
 
 Subsquently, a non-singular family of cosmological models has 
been obtained [4,5]. The only two
equations of state allowed are $\rho=3p$ the radiation model [3] and
$\rho=p$ the stiff fluid model [6]. Further it is also possible to include
heat flux in the non-singular family 
without disturbing the singularity free character [7,5]. All these models
were cylindrically symmetric and having diagnol metric. Mars also found a
non-diagnol stiff fluid model [8]. 

 By enlarging the scope of matter field to include imperfect fluid with
anisotropic pressure and radial heat flux, a family of spherically
symmetric singularity free models has been identified [9,10]. It 
also includes an interesting and novel case of an oscillating singularity
free model [11]. The model oscillates between two regular states without
ever meeting any kind of singularity anywhere. All the models
satisfy the causality and the strong energy conditions with proper fall
off behaviour for all the physical and kinematic parameters. The typical
behaviour is to have low density at $t\rightarrow\pm\infty$ with a peak
at $t=0$. This is also the epoch of contraction turning into expansion and 
vice-versa.

 The Nariai metric was obtained [12] in 1950 when cosmic singularity was
however not much discussed about. It is an Einstein space, which
implies the inflationary equation of state $\rho+p=0$, and remarkably it
is not conformally flat. It is thus not the de Sitter space. It is
geodesically complete and hence is singularity free. It thus automatically 
satisfies the weak energy condition in the limit of the null energy
condition. This is the condition which is believed to hold good always
on the empirical grounds. 

 Let me first demonstrate the non-singular character of the
metric which is given by 
\be
ds^2 = (1+k^2t^2)^{-1}dt^2 - (1+k^2t^2)dz^2 - \frac{1}{k^2}(d\Omega^2)
\ee
where $k$ is a constant and $d\Omega^2 = d\t^2 + \sin^2\t d\varphi^2$. It
could be easily transformed to the form
\be
ds^2 = dt^2 - \cosh^2 kt dz^2 - \frac{1}{k^2}d\omega^2
\ee
by the transformation $kt\rightarrow \sinh kt$. This form clearly exhibits
that spacetime is singularity free and it could be easily verified that it
is geodesically complete. It has the constant density,
\be
\rho = k^2 = -p
\ee  
which could be arbitrarily chosen. It is homogeneous but anisotropic and
has non-zero shear as well as non-zero Weyl curvature. It is thus not
conformally flat.  

 The cosmic dynamics is determined by the Raychaudhuri equation [13] which
reads as follows:
\be
\frac{d{\bf \t}}{ds} = -4\pi(\rho+3p) + 2\omega^2 - 2\sigma^2 -
\frac{1}{3}{\bf{\t}}^2 + \dot u^a_{;a}
\ee
where ${\bf \t}, \sigma, \omega, \dot u^a$ refer respectively to
expansion,
shear, vorticity and acceleration. In this case, the spacetime is
homogeneous and irrotational which means that the acceleration and
vorticity vanish. Since the spacetime does not satisfy the strong
energy condition, the active gravitational charge density,
$\rho_c=\rho+3p$ is in fact negative. In the above equation, shear and
expansion favour contraction while $\rho_c<0$ favours expansion. In the
case of the de Sitter model, shear also vanishes and we are only left
with negative $\rho_c$ and expansion. It is geodesically incomplete as
geodesics terminate at some epoch in the past. The present case 
essentially differs from the de Sitter by the presence of shear. It is thus 
the presence of shear that makes the difference and is responsible for
singularity free character.

 The other difference is that it is not
conformally flat which is again the consequence of non-zero shear. In
absence of acceleration and vorticity, the Weyl curvature is generated by
shear [14]. Intutively, when shear is non-zero, the spacetime is
necessarily anisotropic and consequently it cannot be conformally
flat. The Weyl curvature could however be non-zero for vanishing shear 
spacetime. For instance the 
static Tolman model [15] is shear free but is conformally non-flat. The 
spherical model [8,10] is obtained simply by making it expand.

 Note that ${\bf \t}=k\tanh kt$ and the anisotropy ratio, 
$(\sigma/\t)^2 = 2/3$. It turns out that the Senovilla
[3] and
the spherical [9] models also have the same anisotropy ratio. The expansion 
parameter changes sign at $t=0$, and it is equal to $-k$
for $t\rightarrow-\infty$ and to $k$ for $t\rightarrow\infty$. For
negative $t$ it contracts while for positive $t$ it expands. In 
contrast, it is constant for the de Sitter. This is the distinguishing
feature of singularity free models. There must occur changeover from
contraction to expansion and vice-versa at some finite time.

 For the perfect fluid models obeying the strong energy condition, it is
the acceleration (inhomogeneity) and vorticity oppose while shear
(anisotropy) and expansion favour the collapse in the Raychaudhuri equation. 
In cosmology,
vorticity is not sustainable and hence for avoidance of singualrity
inhomogeneity is necessary. It though turns out that acceleration alone is
never sufficient to check the collapse into singularity. There is no
general result proving this but all the known cases bear it out. I would 
like to conjecture that it has always to be aided by shear and/or heat
flux. Though shear contributes
positively to collapse in the Raychaudhuri equation, it is its  
dynamical action which plays the crucial role of making collapse
incoherent. Consequently it goes on to distracting concentration of large 
mass in small enough a region which is critically required for formation of 
trapped surface leading to singularity. Similar is the case with the heat
flux [9]. It is thus necessary to have shear and/or heat flux
non-zero for singularity free cosmological models. The perfect fluid
singularity free models [3-8] are all both accelerating and shearing while
the imperfect fluid spherical models [9-11] are all accelerating, shearing
and/or also having radial heat flux.

 It is clear that so long as we adhere to the strong energy condition
$\rho_c>0$, there cannot occur a singularity free homogeneous perfect fluid 
model. The present metric indicates that if we instead adhere only to the
weak energy condition (in the null energy condition limit), we could
have homogeneous but anisotropic singularity free model. Since the
gravitational charge density $\rho_c<0$, it would imply
expansion. This property is however shared by the singular de Sitter and
the non-singular Nariai metrics. What is then required for non-singularity
is just to make the expansion incoherent so that the geodesics donot
terminate in the past. This is precisely what the shear does in this case
and renders the model singularity free.

 The Nariai metric belongs to a spacetime which arises out of product
of 2-spaces of constant curvature. In this product spacetime, when the two
curvatures are equal, it is the conformally non-flat Einstein space
described by the Nariai metric. When they are equal but opposite in
sign, it is the conformally flat Bertotti-Robinson metric describing the
uniform electric field [16,17]. In the same framework, we could have a
non-flat metric with one 2-space being flat while the other 2-sphere
having the constant curvature. It would then describe a cloud of string dust 
of uniform energy density [18]. The remarkable feature of this framework
is that it gives examples of the metrics which have unusual
character, opposite to what is generally the case. One associates 
conformal flatness with the Einstein space while conformal non-flatness
with electromagnetic source. Here we have the opposite, the Einstein space
is not conformally flat while the one with the uniform electric field is. 

 From the point of view of application to cosmology, it presents rather a
queer scenario. There is inflationary expansion only in one direction
while the other two remain locked to some constant value determined by the
energy density. The important point to note is that there exists an
inflationary solution other than the de Sitter spacetime. This is again
to indicate that the de Sitter is not unique. It is indeed very
important to recognise the fact that there could exist an alternative to
the de Sitter inflation. Furthermore it has though been shown that it is 
dynamically, like the Einstein universe, not stable [19]. On perturbation, it 
goes over to the de Sitter spacetime. 

 Another remarkable feature is that it has non-zero Weyl curvature which
describes free gravitational field. It would thus be physically very
different from the usual de Sitter inflation. The essential difference
between the two is shear. As mentioned earlier that in this case it is the
shear that produces the Weyl curvature and hence it is directly related to
the free gravitational field. As shear seems to be quite a generic feature
of the singularity free models which suggests that free gravitational
field plays significant role in avoidance of cosmic sularity. That is
singularity free models must necessarily be conformally non-flat. This
would be a necessary condition but however not sufficient. 

 Thus the difference between the Nariai and the de Sitter spacetimes is shear. 
It is interesting to note that on perturbation the former tends to the latter. 
That means perturbations tend to kill shear and thereby take the Nariai 
metric to the shear free de Sitter spacetime. All this exhibits an interesting 
interplay of shear, isotropy, Weyl curvature and occurrence of singularity.

 If we wish to consider a slightly more general case than that of the
perfectly homogeneous and isotropic universe, what is it that could be
included without seriously disturbing the overall scenario? Obviously it
would be shear and anisotropy, yet retaining homogeneity. The present case
would then become pertinent as it gives inflation with shear. For
this we have to go off spherical symmetry, because the de Sitter is
the unique spherically symmetric Einstein space solution. The present
spacetime does contain a sphere of constant radius. It could thus
perhaps be considered anisotropic generalization of the de Sitter
space. We would like to say that it falls within the pertinent
generalization space of the standard inflationary model. It would be 
interesting to examine whether it could provide a tenable alternative 
inflationary scenario. 

 At present, we
just wish to point out that the Nariai metric is the first example of the
singularity free models satisfying the weak/null energy condition. Further, if
the only weak/null energy condition is to be adhered to, then it is possible 
to have homogeneous singularity free model and the Nariai metric is an 
example of that.

\end{document}